\documentclass{Interspeech2024}

\usepackage{dcolumn}
\usepackage{multirow}
\usepackage{hyperref}
\usepackage[table]{xcolor}
\usepackage{fontawesome5}
\usepackage{svg}
\usepackage{subcaption}
\usepackage{ amssymb }
\usepackage{oplotsymbl}

\interspeechcameraready

\title{On the calibration of powerset speaker diarization models}
\name{Alexis}{Plaquet}
\name{Hervé}{Bredin}
\address{
  IRIT, Université de Toulouse, CNRS, Toulouse INP, UT3, Toulouse, France
}
\email{firstname.lastname@irit.fr}


\begin{document}

\maketitle

\begin{abstract}
    End-to-end neural diarization models have usually relied on a multilabel-classification formulation of the speaker diarization problem. Recently, we proposed a powerset multiclass formulation that has beaten the state-of-the-art on multiple datasets. In this paper, we propose to study the calibration of a powerset speaker diarization model, and explore some of its uses.
    We study the calibration in-domain, as well as out-of-domain, and explore the data in low-confidence regions. The reliability of model confidence is then tested in practice: we use the confidence of the pretrained model to selectively create training and validation subsets out of unannotated data, and compare this to random selection.
    We find that top-label confidence can be used to reliably predict high-error regions.  Moreover, training on low-confidence regions provides a better calibrated model, and validating on low-confidence regions can be more annotation-efficient than random regions.
\end{abstract}


\section{Introduction}
The speaker diarization task can be defined as taking an audio excerpt and answering the question ``who spoke when?'', without concern for the exact identities of the speakers. Solving this task provides the exact beginning and end of each speaker turn, which proves very useful when combined with other tasks output such as transcribed text.

Classifiers usually provide some notion of ``confidence'' along with the predicted output. In an ideal world, confidence would always be linked to the epistemic or aleatoric uncertainty. However, deep learning classifiers are famously overconfident and predict high probabilities for unknown classes and classes where the model is wrong \cite{guo_cal_2017}. Despite these limitations, model output probabilities are still one of the only tools available to estimate uncertainty contained in the predictions of deep learning models. This has led to a number of different usages: out-of-domain detection \cite{odin2017}, semi-supervised learning \cite{pseudolabel}, or active learning.

Research on the calibration of End-to-end Neural Diarization (EEND) models and its application is lacking. The goal of this paper is to study the calibration of the powerset speaker diarization model proposed in \cite{plaquet23}, and to use model confidence to select data of interest for training and validation purposes. We study in detail the calibration of the model on in-domain and out-of-domain data, and observe what kind of data is represented in low confidence predictions. Moreover, we study the selection of low-data validation and training set with confidence-based strategies.


\section{Model calibration}



\begin{figure}[!b]
    \centering
    \includegraphics[width=0.76\linewidth]{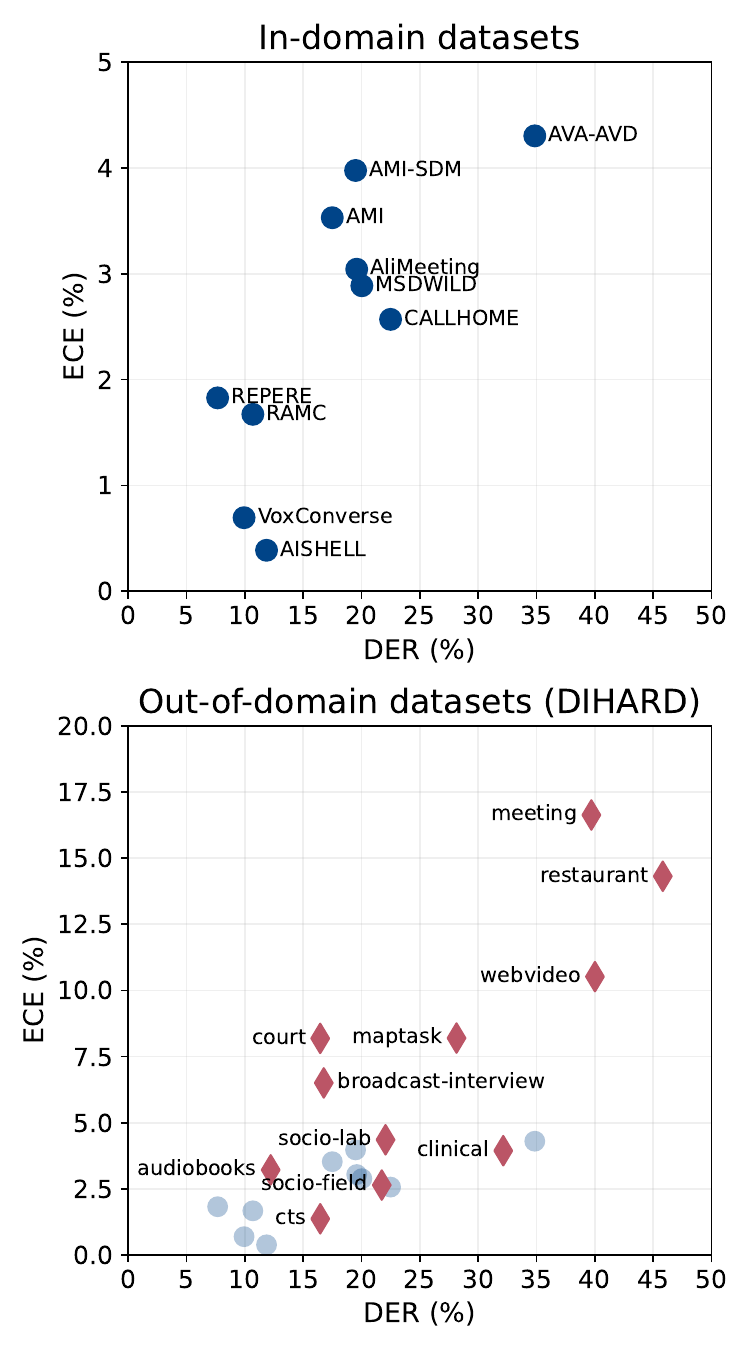}
    \caption{Calibration error as a function of the DER of the \textit{powerset} segmentation model. In-domain datasets (top figure) are plotted with blue circles, out-of-domain datasets (bottom figure) are plotted with red diamonds. To give a frame of reference, the blue circles of in-domain datasets are also overlayed in transparency on the bottom figure.}
    \label{fig:summary-calibration}
\end{figure}

\begin{figure}[ht]
    \centering
    \includegraphics[width=\linewidth]{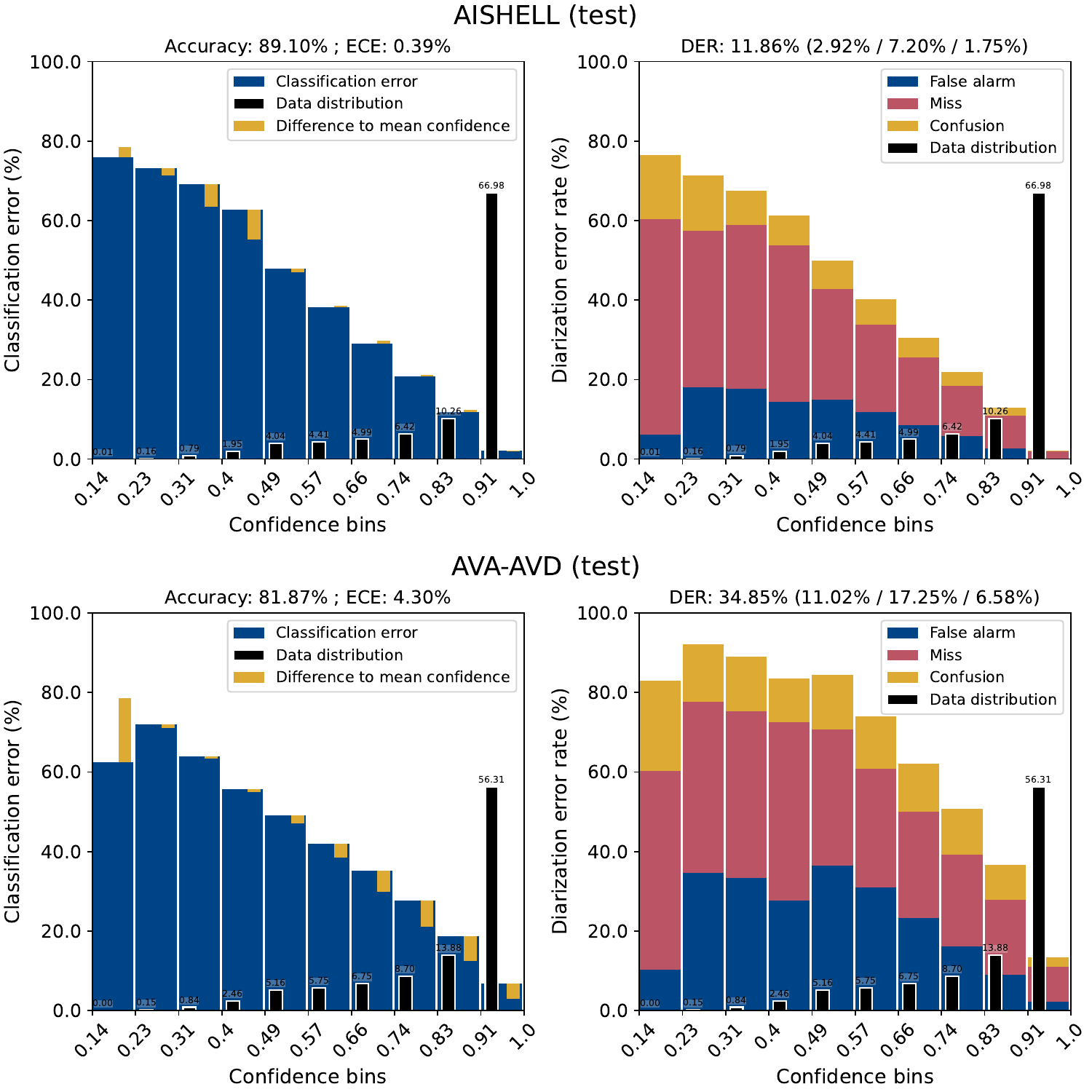}
    \caption{Best and worst in-domain calibration, measured by ECE. The left column is a classical reliability diagram, a perfect ECE would mean no ``difference to mean confidence`` in every bin, resulting in a diagonal plot. The right column is the same plot but with DER instead of classification error.}
    \label{fig:indomain-calibration}
\end{figure}

\begin{figure}[ht]
    \centering

    \includegraphics[width=1\linewidth]{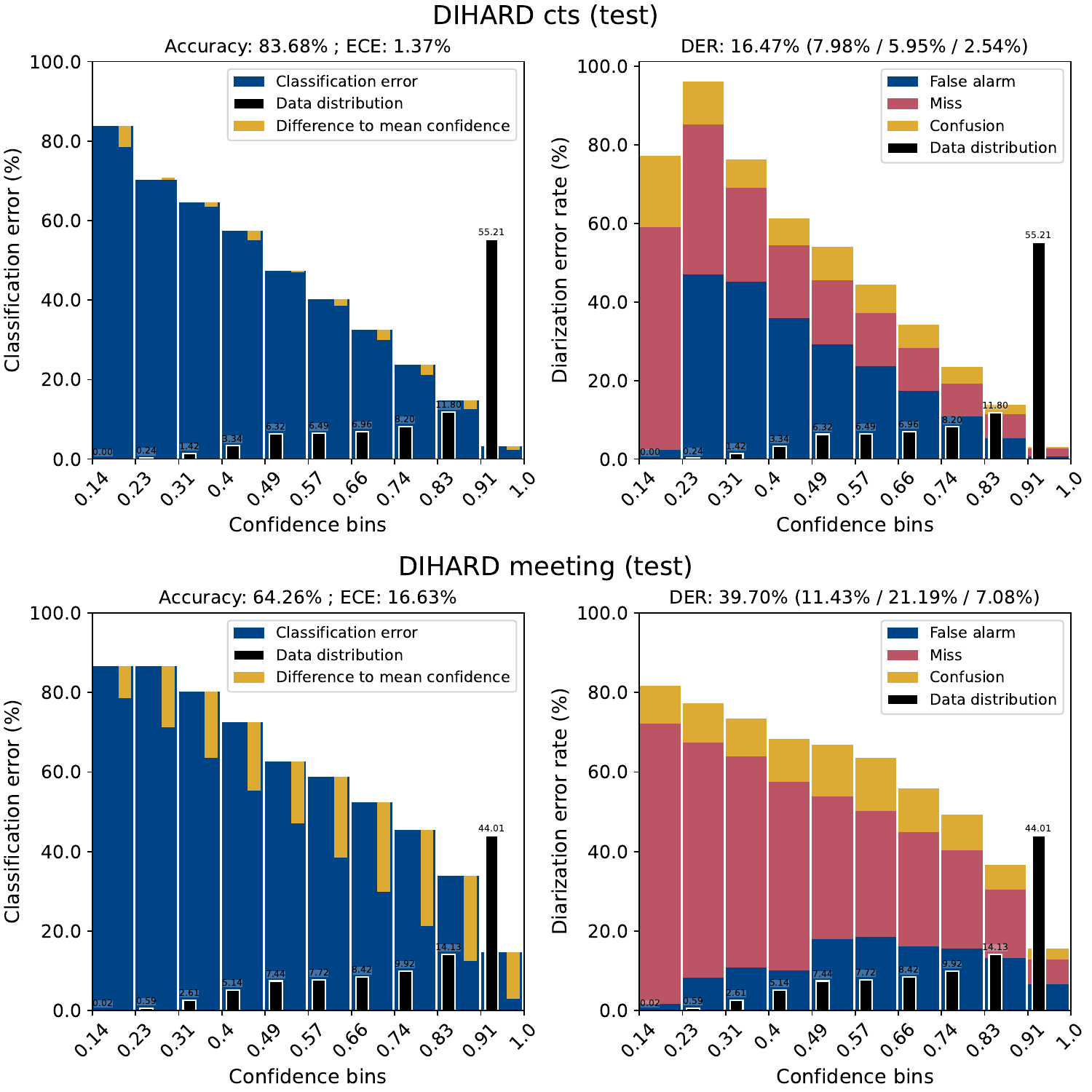}
    \caption{Reliability diagram and binwise DER distributions for the best and worst calibrated domains in DIHARD.}
    \label{fig:outofdomain-calibration}
\end{figure}

\begin{figure}[ht]
    \centering
    \includegraphics[width=1\linewidth]{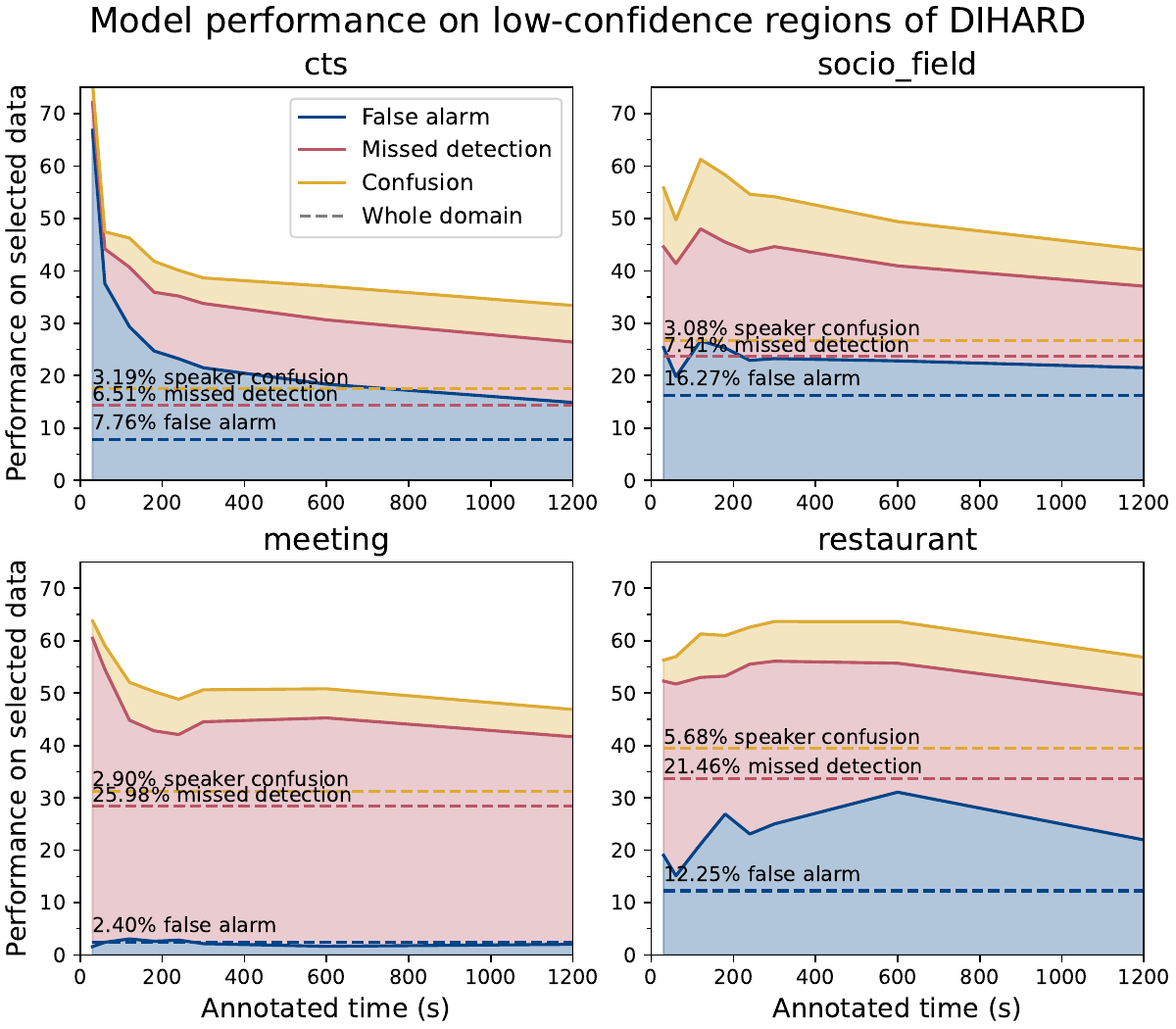}
    \caption{Composition of the diarization error rate when sampling 5 seconds chunks, lowest confidence chunks are selected first. The dashed lines show the composition of the DER on the whole test set.}
    \label{fig:der-distribution}
\end{figure}

\begin{figure}[ht]
    \centering
    \includegraphics[width=1\linewidth]{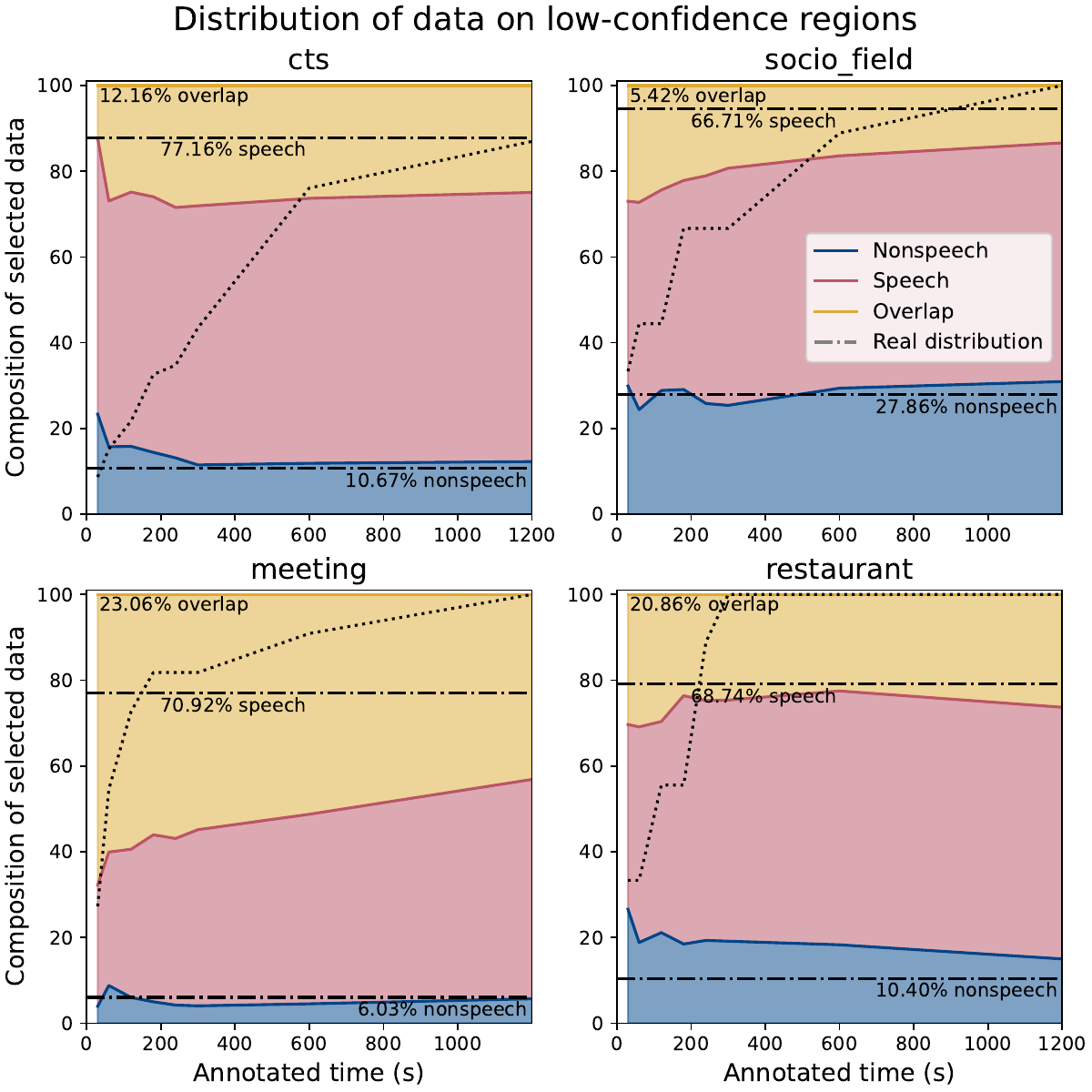}
    \caption{Composition of the data categorized in nonspeech, speech and overlap, when sampling 5 seconds chunks like in \autoref{fig:der-distribution}. Dashed lines show the average distribution on the whole test set.}
    \label{fig:data-distribution}
\end{figure}

In a multiclass setting, a model is deemed ``top-label-calibrated'' if the maximum of its output scores (the score of the predicted class) is equal to the probability of it being the correct prediction. For example, a top-label-calibrated model outputting probabilities $(c_1,c_2,c_3)=(0.7,0.1,0.2)$ has a 70\% chance of being correct about $c_1$ being the correct class. Stricter definitions of calibration exist, such as classwise-calibrated~\cite{classwise_calibration} and jointly-calibrated~\cite{beyond_calibration}, but they are out of the scope of this paper.

There is no easy way to guarantee a degree of model calibration in deep learning. Models are not designed to achieve top-label-calibration, and tend to be naturally overconfident~\cite{guo_cal_2017}. There are two main ways to approach top-label-calibration: encourage better calibration during training, and post-hoc calibration methods~\cite{SilvaFilho2023}. Training-time calibration methods are diverse and encompass regularization methods~\cite{label_smoothing} or loss modifications~\cite{logitnorm}. Simple post-hoc calibration methods work well~\cite{guo_cal_2017}, but are very sensitive to data shift: calibrating on a domain will only help on data that is part of, or very close to this domain \cite{ovadia19_trust_uncertainty}.
This means that it is extremely difficult to design a model that is calibrated on any data it encounters: ``top-label-calibration'' is not free and has huge annotation cost.

We study the ``top-label-calibration'' of the local speaker diarization model proposed in~\cite{plaquet23}. To do so, we trained a standard powerset PyanNet model with classes for at most 2 simultaneous speakers, and up to 3 distinct speakers in 5s chunks. The model is trained until convergence after 89 hours of training on a single NVIDIA V100 GPU. We rely on a compound training dataset made of the concatenation of the training subsets of AISHELL-4~\cite{AISHELL}, AliMeeting~\cite{AliMeeting}, AMI~\cite{AMI}, AVA-AVD~\cite{AVA-AVD}, CallHome~\cite{CALLHOME}, Displace~\cite{DISPLACE}, Ego4D~\cite{Ego4D}, MSDWild~\cite{msdwild}, MagicData-RAMC~\cite{RAMC}, REPERE~\cite{REPERE}, and VoxConverse~\cite{VoxConverse}. Any of the test subsets from the compound dataset is considered as ``in-domain''. We use the 11 distinct sub-domains of DIHARD 3~\cite{DIHARD} dataset as ``out-of-domain'' datasets (\textit{i.e.} that have never been seen during training).

In the spirit of reproducible research, all relevant model checkpoints, metrics, metadata and code are available at {\scriptsize\url{github.com/FrenchKrab/IS2024-powerset-calibration}}.

\subsection{Metrics}

In-domain performance is assessed with two metrics: diarization error rate (DER) and expected calibration error (ECE). 

To compute ECE, model predictions are grouped by their confidence into different bins. As the powerset model outputs softmax probabilities, we define the confidence as the probability of the predicted class. ECE is defined as:
\begin{equation*}
    \text{ECE} = \sum_{i=0}^B \text{prop}(b_i) \cdot | \text{acc}(b_i) - \text{conf}(b_i) |
\end{equation*}
Where $\text{prop}(b_i)$ is the proportion of predictions in bin $b_i$, and $\text{acc}(b_i)$ and $\text{conf}(b_i)$ the average accuracy and confidence in $b_i$.
Multiple binning schemes and distances $| \text{acc}(b_i) - \text{conf}(b_i) |$ can be used. In our experiments and figures, we use $N=10$ bins uniformly distributed in $[\frac{1}{\text{class count}}; 1]$, and the L1 distance. We experimented with Adaptive ECE (where all bins contain the same number of samples), and varied the bin size from 10 to 20, but we did not find any meaningful differences and hence we do not report the different variants.

The DER is a standard speaker diarization metric defined as
\begin{equation*}
    \text{DER} = \frac{\text{False alarm} + \text{Missed detection} + \text{Speaker confusion}}{\text{Total speech}}
\end{equation*}
It is commonly expressed in percentages but can go over 100\% as false alarm can exceed the total duration of speech.

Since our focus is the local segmentation model working on 5 seconds chunks only, we disregard the usual subsequent steps of the diarization pipeline (embedding extraction and clustering)~\cite{pyannote21}. Metrics are directly computed on the (sliding) outputs of the local segmentation model after each window has been aligned with the reference. We call this ``local DER'' and it should not be compared to the values of DER that are usually reported in the literature.

\subsection{In-domain and out-of-domain calibration}

The ECE and DER on the test subsets of all datasets are shown in~\autoref{fig:summary-calibration}. We observe some correlation between DER and ECE: domains with higher DER tend to have higher ECE as well. On in-domain datasets, the ECE goes up to 4.3\% on AVA-AVD with a DER of 35\%, the worst performing dataset DER-wise and ECE-wise. However, most out-of-domain datasets are badly calibrated, only 4 out of 11 are under 4.3\% ECE. And, unsurprisingly, the DER is also worse on out-of-domain datasets.

The left row of \autoref{fig:indomain-calibration} shows reliability diagrams on a couple of in-domain datasets.
On AISHELL, the model has nearly perfect calibration (the best out of all tested in-domain datasets): the model confidence matches its average accuracy very well. Even though AVA-AVD is the in-domain dataset where the model has the worst calibration, it is still fairly low with only $4.30\%$. However, we can see a trend where the mis-calibration comes mainly from model overconfidence: the model makes too many errors in high confidence bins. 

\autoref{fig:outofdomain-calibration} shows the same plot for out-of-domain datasets. The ``meeting'' plot displays how the model is overconfident in all bins, which is usually the expected pattern in deep neural networks. We observe a very high ``missed detection'' rate, which suggests that the model is not sensitive enough to detect the speech and fails to reflect any uncertainty in its output probabilities. The distribution of errors (false alarm, missed detection, speaker confusion) does not depend on the ECE but rather on the mismatch between the domain and the pretrained model's training data.

\subsection{Analysis of low-confidence regions}
\label{sec:lowconfidence_regions}

Previous figures provide insights into \textit{framewise} calibration and confidence. However one of the main challenges of speaker diarization is the temporal aspect of the prediction. While in image classification it makes sense to discard or select individual data samples, in speaker diarization it does not always make sense to only keep or select individual frames of data. This is especially true in use-cases involving human annotators, as they need the preceding audio context to make sense of a specific frame. Consequently, in the following figures all selected data is made of continuous chunks of at least 7.5 seconds. We select in priority regions where the average confidence is the lowest (i.e ``low-confidence regions'') and study model performance and class distribution on them.

In \autoref{fig:der-distribution}, we look at the two best (CTS, Socio-field) and worst (Meeting, Restaurant) calibrated domains. The figure shows the diarization error rate when selecting $x$ seconds of data.
In both CTS and socio-field, the DER in low-confidence regions for $x\in[0,1200]$ seconds is significantly higher than the DER computed on the whole domain. In Meeting and Restaurant, this difference is less significant.
Nonetheless, even in the worst calibrated dataset, DER on low-confidence regions stays significantly higher than the average on the domain. This observation is encouraging, as it appears that even when the model is badly calibrated on a domain, low-confidence still strongly correlate with low performance.

\autoref{fig:data-distribution} shows input data distribution in the same datasets on the same low-confidence regions. We divide the data into three categories: ``nonspeech'' when no speaker is active, ``speech'' when only one speaker is active, and ``overlap''' when at least two speakers are active simultaneously. In both well- and badly-calibrated datasets, we observe that low-confidence data contains significantly more overlap than the rest of the classes. This can be expected as overlapped speech is difficult, and remains one of the main sources of error in speaker diarization.

\section{Annotation-efficient domain adaptation}

Although speaker diarization models are getting better and better, applying a pretrained state-of-the-art speaker diarization model on unseen data might not output what we expect. This could be because the model has not yet seen or generalized to this specific kind of data: acoustic conditions, speech type, language. Or it could simply be because annotation standards vary a lot between datasets~\cite{pyannote21}, and the ones learnt by the model might not fit the needs of the user. In any case, if one wants optimal performance on a new dataset, one will need to specify what is expected from the model. This usually means annotating (at least some of) the data, which is a very costly process, requiring up to dozens of person-hours to label a few hours of audio.

In this part, we borrow from active learning the idea of focusing the human annotation effort to low-confidence regions of the data. This choice is motivated by the results obtained in \autoref{sec:lowconfidence_regions} where we observe a high DER on low-confidence regions. 

We treat DIHARD 3 eleven distinct datasets as ``unlabeled'' for these experiments (artificially withholding annotations). We simulate the human annotation process with an ``oracle labeler'' which provides the withheld annotations when requested. Our goal is to improve the DER while using as little oracle annotations as possible.

\subsection{Finding a minimal training subset}

\begin{figure}[!b]
    \centering
    \includegraphics[width=0.98\linewidth]{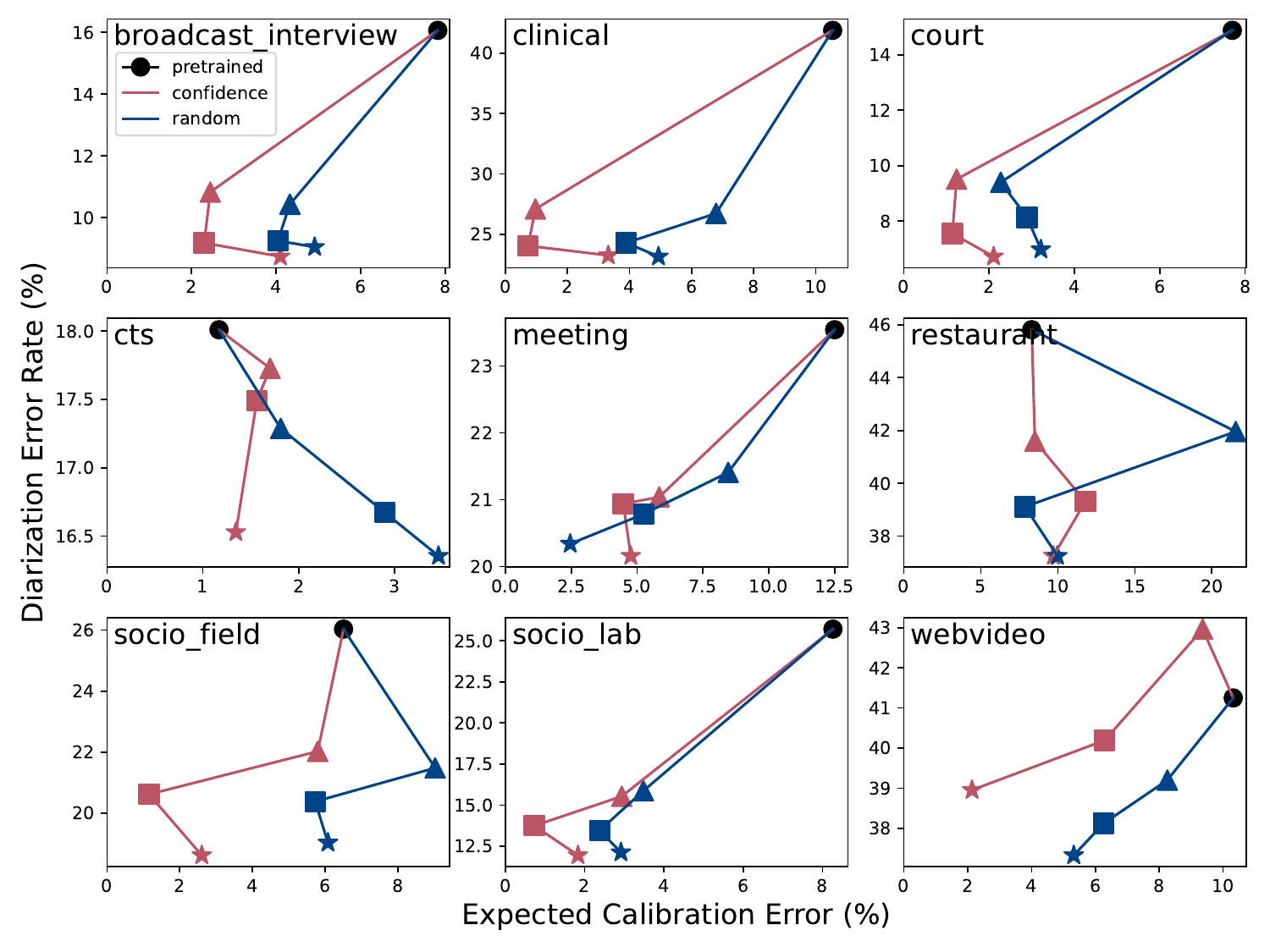}
    \caption{Evolution of DER and ECE when varying the size of the training dataset. Each marker corresponds to a dataset duration: $\blacktriangle$=30s, $\blacksquare$=2min30s, $\bigstar$=20min.}
    \label{fig:training-summary}
\end{figure}

In active learning the learning process is usually iterative, but here we limit the research to a single iteration: we select the relevant data to annotate, label the (withheld) selected regions, retrain the model on this new data, and finally evaluate the model performance. The selected regions are 7.5 seconds long to mirror the process described in \autoref{sec:lowconfidence_regions}.
We test two ways to select the regions to label: 
\begin{itemize}
    \item Random: the data used to train the model is selected at random (our baseline).
    \item Worst confidence: the regions where the average confidence of the model is the lowest are selected (like in \autoref{sec:lowconfidence_regions}).
\end{itemize}

We can make some interesting observations on the results that are summarized in \autoref{fig:training-summary}. First, for all domains but webvideo, 30 seconds of training data is enough to significantly improve the DER. The improvement can be quite important in-domains like socio-lab or court. Webvideo behaves this way probably because it is not a homogeneous domain, but a collection of heterogeneous YouTube videos, hence the need for more data. More importantly, we can observe that the querying strategy does not have a strong impact on the DER either way. At equal annotation budget, we can expect a similar DER. However, the confidence-based selection obtains better ECE on almost all domains.

\subsection{Finding a minimal validation subset}

One limitation of retraining the model for domain adaptation is the need for a validation set. Without a validation set, we might choose a checkpoint where the model has overfitted, which is very likely on low amounts of data.

However, we also need the validation set to be as small as possible if we want the model retraining to effectively be annotation-efficient (as the last subsection used oracle validation). We need to find the smallest validation subset that selects the same best checkpoint as the full validation set.

This task proves to be difficult, even if we allow suboptimal selection of checkpoint (i.e. we allow a small relative difference in DER between the best checkpoint selected from a small validation set, and the real best checkpoint selected from the full validation set).

To evaluate it, we use a fixed set of checkpoints, compute the DER on validation subsets of various lengths, using random and low-confidence selection, and observe how good they approximate the full set. We estimate that between 2 and 5 minutes of data are required to reliably select a checkpoint with less than 10\% of relative difference in DER to the best checkpoint. More data is needed if we want to approach more closely the selection of the full validation set. Interestingly, random selection of regions yields a better minimal validation set at low annotation budget (under 5 minutes), but is outclassed by the selection of low-confidence regions when the budget increases.

\section{Conclusion}

In this paper, we studied the calibration and performance of the powerset speaker diarization model on 12 datasets seen during training, as well as the 11 domains composing DIHARD 3. We found that the model is well calibrated on in-domain datasets, while calibration on out-of-domain datasets is generally worse. Despite this, we observed that diarization error rate on predicted low-confidence regions is always significantly higher than the average on the dataset.

We then simulated the annotation of low-confidence regions on out-of-domain datasets to constitute small training sets. We observed that such sets offer no significant advantage or disadvantage to random region selection in terms of DER, but prove to be better calibrated. Selection of a minimal validation set proves to be a difficult task, but selection of low-confidence regions seems to improve its efficiency given a high enough annotation budget.

These results lead us to believe that top-label confidence can be reliably used to find regions of the data where powerset speaker diarization model performs badly. Uses include out-of-domain detection, semi-supervised learning, and especially active learning. The improvement in calibration after domain adaptation is very encouraging and lead us to believe that active learning with iterative retraining and selection of new low-confidence regions using the better calibrated model might take full advantage of this property.


\section{Acknowledgements}

This work was granted access to the HPC resources of IDRIS under the allocation AD011013477R1 made by GENCI.
The research described in this paper is partly supported by the Agence de l'Innovation Défense under the grant number 2022 65 0079.

\newpage

\bibliographystyle{IEEEtran}
\bibliography{mybib}

\end{document}